# Growth of strontium ruthenate films by hybrid molecular beam epitaxy

**Patrick Marshall[a)], Honggyu Kim, Kaveh Ahadi, and Susanne Stemmer[b)]**

Materials Department, University of California, Santa Barbara, CA 93106-5050, USA

[a)] Electronic mail: pmarshall@mrl.ucsb.edu

[b)] Electronic mail: stemmer@mrl.ucsb.edu

**Abstract**

We report on the growth of epitaxial $Sr_2RuO_4$ films using a hybrid molecular beam epitaxy approach in which a volatile precursor containing $RuO_4$ is used to supply ruthenium and oxygen. The use of the precursor overcomes a number of issues encountered in traditional MBE that uses elemental metal sources. Phase-pure, epitaxial thin films of $Sr_2RuO_4$ are obtained. At high substrate temperatures, growth proceeds in a layer-by-layer mode with intensity oscillations observed in reflection high-energy electron diffraction. Films are of high structural quality, as documented by x-ray diffraction, atomic force microscopy, and transmission electron microscopy. The method should be suitable for the growth of other complex oxides containing ruthenium, opening up opportunities to investigate thin films that host rich exotic ground states.

The discovery of unusual ground states in the layered ruthenates, such as p-wave superconductivity [1,2] and nematic order [3], has led to significant interest in the growth of high-quality thin films of these materials. For example, $Sr_2RuO_4$ is an unconventional, spin triplet superconductor [4], whose superconducting transition temperature is highly sensitive to strain [5]. Thin films of $Sr_2RuO_4$ would allow for advancing the understanding of unconventional forms of superconductivity through the application of epitaxial strain [6] and are needed for applications in topological and hybrid superconducting devices. A remarkable success of pulsed laser deposition (PLD) is the growth of superconducting $Sr_2RuO_4$ films [7,8], because even minute defect concentrations destroy the superconductivity completely [9].

Molecular beam epitaxy (MBE) has been key in the development of high quality semiconductor thin film heterostructures [10] and has produced high performance complex oxide thin films [11]. MBE of the ruthenates is, however, challenging because of the extremely low volatility of Ru [12], requiring the use of electron beam assisted evaporation [13,14]. Electron beam sources suffer from inherent flux variations and require in-situ monitoring and real time feedback to continuously adjust the flux [14], which makes stoichiometry control difficult. Other challenges include practical limits on the oxygen pressure/fluxes that can be used and flux instabilities in the presence of oxygen [15].

An alternative approach is hybrid MBE, which uses a combination of solid and gas sources to supply the metals [16,17]. In particular, metalorganic and other gaseous sources that have much higher vapor pressures and already contain metals bonded to oxygen, and whose fluxes can be precisely controlled, have been shown to successfully addresses all of the challenges in MBE of perovskite titanates [17-19], zirconates [20], and vanadates [21], resulting in high quality thin films having record properties [22,23].

Here, we demonstrate the growth of epitaxial $Sr_2RuO_4$ films using hybrid MBE. Strontium is evaporated from an elemental metal source while a volatile ruthenium tetroxide ($RuO_4$)-containing precursor is used to supply pre-oxidized Ru [24]. The $RuO_4$ precursor has been used to synthesize Ru, $RuO_2$, and $SrRuO_3$ films [24-26], but not yet in MBE. The resulting films are shown to be of excellent structural quality. The approach should be suitable for other ruthenate systems as well, thus opening up a viable MBE approach for thin films of a very exciting class of materials.

$Sr_2RuO_4$ films were grown on (001) $(LaAlO_3)_{0.3}(Sr_2AlTaO_6)_{0.7}$ (LSAT) substrates in an oxide MBE system (GEN 930, Veeco Instruments). The in-plane lattice parameters of $Sr_2RuO_4$ and LSAT are (nearly) perfectly matched (3.87 Å [27,28]), making LSAT an excellent substrate for the epitaxial growth of strain-free $Sr_2RuO_4$. Substrates were etched in a 3:1 solution of $HCl:HNO_3$ and backed with 350 nm Ta to facilitate the radiative heating from the substrate heater during growth. Films were grown by co-deposition. Strontium was supplied from a high-purity elemental source contained in a low-temperature effusion cell. Ruthenium was supplied in the form of the $RuO_4$-containing precursor [24]. This precursor was packed with fluorinated ethers to stabilize the highly reactive $RuO_4$ molecules, which comprised about 10% by weight of the mixture. The precursor cylinder was at room temperature and no carrier gas was used, as the vapor pressure of $RuO_4$ is sufficiently high for delivery to the growth chamber. The precursor mixture was delivered to the growth chamber via a gas inlet system using a capacitive manometer and a linear leak valve to monitor and control the flow rate. It was introduced to the growth chamber through a gas injector heated to 50 °C. Additional oxygen was supplied using a plasma source operated at a power of 300 W with a background pressure of ~ $3\times10^{-6}$ Torr during growth. Film growth and structure were monitored *in-situ* using reflection high-energy electron diffraction (RHEED), *ex-situ* via x-ray diffraction (XRD), atomic-force microscopy (AFM), and transmission electron microscopy (TEM). Room temperature electrical measurement were performed in four point probe (Van der Pauw) geometry using Au/Ti ohmic contacts.

The film stoichiometry was varied by changing the Sr flux via the effusion cell temperature and keeping all the other fluxes constant. Similar to hybrid MBE of $SrTiO_3$ [29], post-growth RHEED patterns (Fig. 1) were found to be sensitive to the film stoichiometry. For Sr-deficient films, the RHEED pattern was noticeable spotty, especially along the [110] direction, while chevron patterns became visible in Sr-rich films. The streaky post-growth RHEED patterns along the [100] and [110] directions for the stoichiometric films indicated a smooth film surface.

The stoichiometry was also evident in the electrical properties, as the room temperature resistivities of the Sr-deficient, stoichiometric, and Sr-rich films were 509 $\mu\Omega$cm, 230 $\mu\Omega$cm, and 643 $\mu\Omega$cm, respectively. In contrast, the out-of-plane lattice parameter (12.74 Å) was found to be independent of stoichiometry. Thus, the best indicators of stoichiometric $Sr_2RuO_4$ films were high intensity and streakiness of the RHEED pattern, prominent thickness fringes in XRD, and low resistivity.

In addition to the flux ratios during growth, high substrate temperatures were essential. Films grown at substrate temperatures of less than 920 °C (thermocouple reading) showed no oscillations in RHEED during growth and were highly resistive. The use of higher substrate temperatures during growth also resulted in a substantial decrease in the room temperature resistivity, from 1100 $\mu\Omega$cm for films grown at 920 °C to 230 $\mu\Omega$cm for films grown at 950 °C. Furthermore, the highest substrate temperatures (950 °C) used in this study were found to promote layer-by-layer growth (see below). All results shown here are from films grown at 950 °C.

RHEED intensity oscillations, shown in Fig. 2(a), indicated layer-by-layer growth mode. This growth mode allows for excellent control over layer thickness and is a signature of sufficiently high adatom mobility on the growing film surface. The initial behavior of the RHEED intensity may be a result of the change in symmetry from the cubic substrate to the tetragonal film, as described previously for PLD-grown films [7]. The period of the intensity oscillations corresponded to a growth rate of 16 nm/hr, with two intensity oscillations per $Sr_2RuO_4$ unit cell, as has also been found by others [30]. AFM [Fig. 2(b)] confirmed the smooth film surface with features that are on order of the unit cell height (12.74 Å).

A wide angle XRD scan is shown in Fig. 3(a). The 002, 004, 006, 008, and 0010 Bragg reflections of $Sr_2RuO_4$ are all present. The lattice spacing is consistent with an out-of-plane lattice constant of 12.74 Å, matching the known value for the tetragonal unit cell of $Sr_2RuO_4$ (space group I4/mmm [31]). Thus, the films are phase-pure $Sr_2RuO_4$. This is significant, as mixed-phase films, containing perovskite $SrRuO_3$ and the $n = 2$ Ruddlesden-Popper phase $Sr_3Ru_2O_7$, are also possible [32]. Figure 3(b) shows a high-resolution scan around the 006 reflection of $Sr_2RuO_4$ and the 002 reflection of the LSAT substrate, respectively. The presence of prominent thickness fringes indicates very smooth interfaces and surfaces. The spacing of the fringes corresponds to a film thickness of 29 nm, in good agreement with the thickness derived from the period of the RHEED oscillations. The rocking curve full width at half max of the 006 $Sr_2RuO_4$ peak was 0.075°, confirming a high crystalline quality of the film [see Fig. 3(c)]. Epitaxial growth of phase-pure $Sr_2RuO_4$ was also confirmed by cross-section high-angle annular dark-field imaging (HAADF) in scanning TEM (STEM), as shown in Fig. 4. The films consist only of the $n = 1$ Ruddlesden-Popper layered crystal structure of $Sr_2RuO_4$. The images show abrupt interfaces and no extended defects.

To summarize, we have demonstrated the growth of phase pure thin films of $Sr_2RuO_4$ with high structural quality using a hybrid molecular beam epitaxy approach that utilizes a highly volatile precursor containing reactive $RuO_4$ molecules. The method allows for control over the stoichiometry and layer-by-layer growth. Future studies will focus on the characterization of the electrical properties as a function of temperature. The excellent structural quality of the $Sr_2RuO_4$ film points towards this precursor as a promising route to the development of other materials in the strontium ruthenate family.

The work was supported by a MURI funded by the U.S. Army Research Office (Grant No. W911NF-16-1-0361) and by FAME, one of six centers of STARnet, a Semiconductor Research Corporation program sponsored by MARCO and DARPA. The microscopy work was supported by the U.S. Department of Energy (Grant No. DEFG02-02ER45994).

**Figure Captions**

**Figure 1:** Evolution of RHEED patterns with varying stoichiometry of the $Sr_2RuO_4$ films. From left to right, the strontium effusion cell temperature was set at 480 °C (Sr-deficient), 488 °C (stoichiometric), and 498 °C (Sr-rich), with the $RuO_4$-containing flux held constant. The substrate temperature was maintained at 950 °C for all three films.

**Figure 2:** (a) RHEED intensity oscillations at the beginning of growth, indicating that growth proceeds in a layer-by-layer mode. (b) 2 μm×2 μm AFM image of the $Sr_2RuO_4$ film surface. The surface is atomically flat with no distinct features.

**Figure 3:** (a) Wide-angle XRD scan of $Sr_2RuO_4$ grown on LSAT. From left to right, the 002, 004, 006, 008, and 00<u>10</u> Bragg reflections of $Sr_2RuO_4$ are present (red circles), corresponding to an out-of-plane (growth direction) lattice constant of 12.74 Å. The peaks from the LSAT substrate (green squares) correspond to the 001, 002, and 003 reflections. (b) High resolution XRD scan showing the 006 reflection of $Sr_2RuO_4$ and the 002 reflection of the LSAT substrate. The prominent thickness fringes indicate a high quality film with smooth interfaces, with the fringe spacing corresponding to a $Sr_2RuO_4$ film thickness of 29 nm. (c) Rocking curve of the 006 peak of a stoichiometric $Sr_2RuO_4$ film. The full width at half maximum is 0.075°.

**Figure 4:** Cross-section HAADF-STEM images of a $Sr_2RuO_4$ film grown on LSAT. (a) Large field of view showing epitaxial growth with smooth interfaces and no extended defects. (b) Higher magnification view confirms the tetragonal $Sr_2RuO_4$ crystal structure with the *c*-axis parallel to the growth direction.

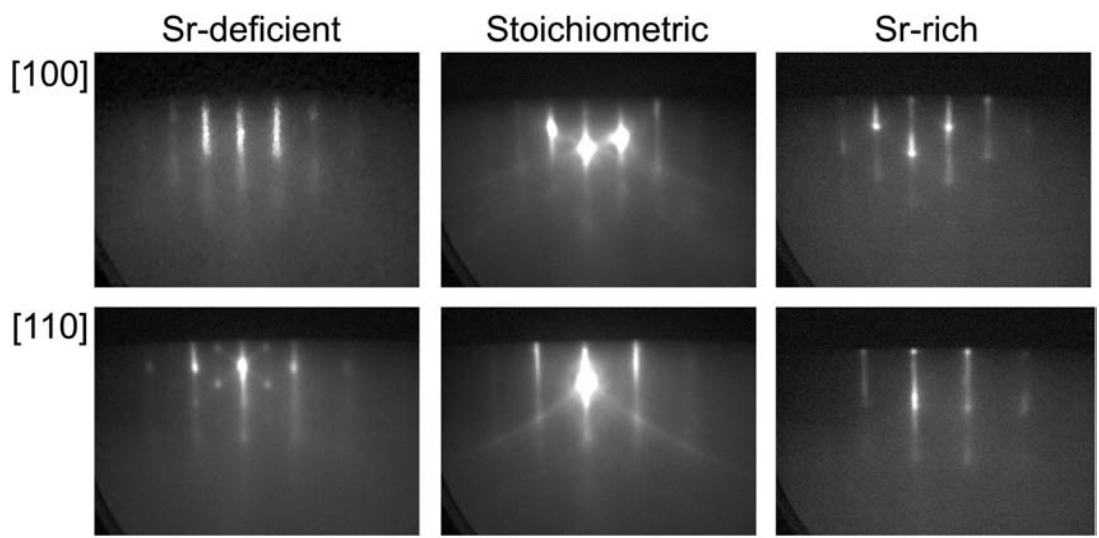

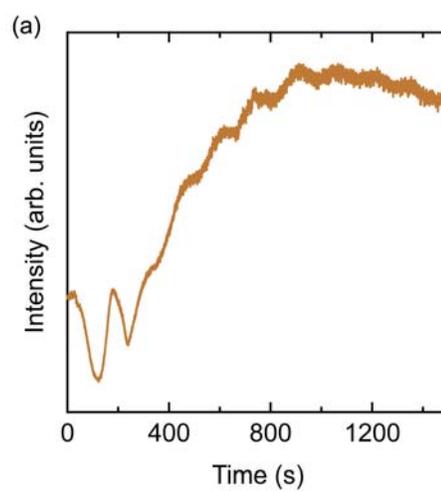

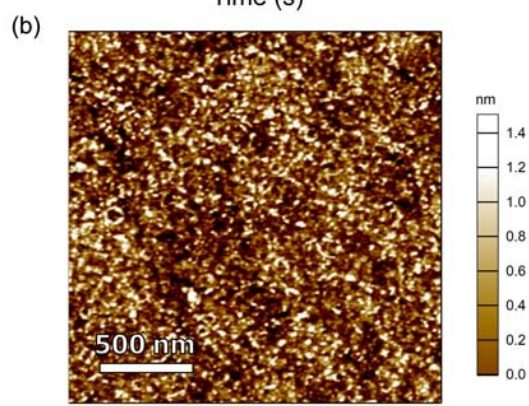

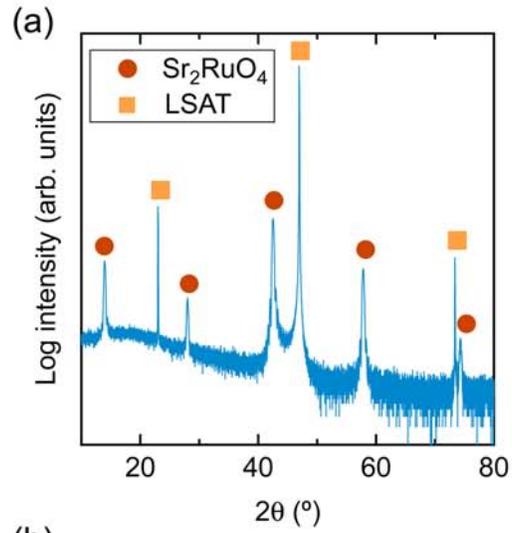
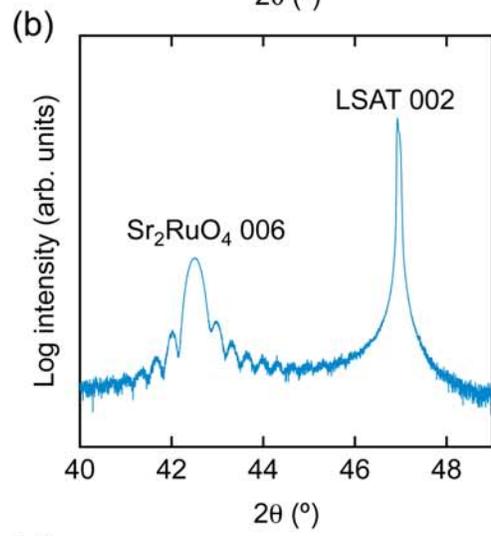
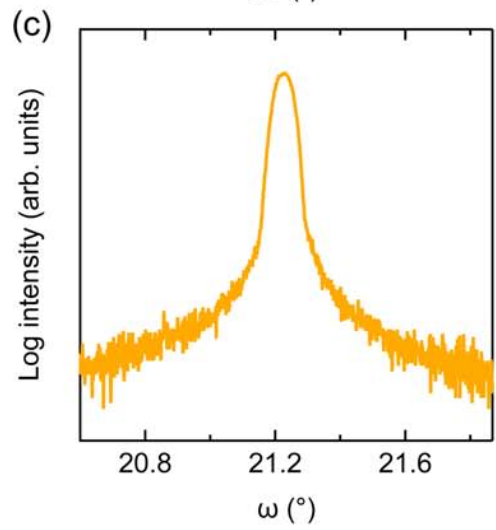

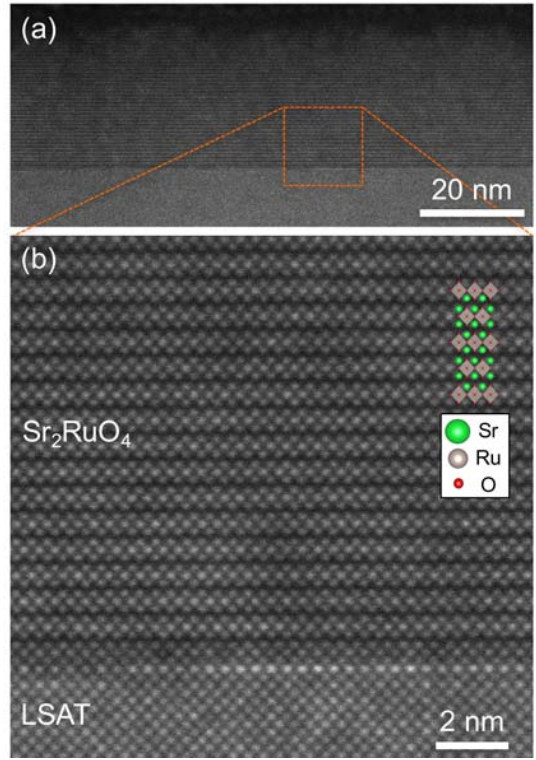